\documentclass[aps,prl,reprint,amsmath,amssymb,showpacs,nobibnotes]{revtex4-1}
\usepackage{graphicx}
\usepackage{hyperref} 
\usepackage{dsfont}    
\usepackage{color}

\newcommand\trick[1]{}

\newcommand{\be}{\begin{equation}}
\newcommand{\ee}{\end{equation}}
\newcommand{\eq}[1]{(\ref{#1})}

\begin{document}
\hfill{ NCTS-TH/1701} 
\title{A New Proposal for Holographic BCFT}

\author{Rong-Xin Miao${}^{1}$ }
\email{miaorongxin.physics@gmail.com}
\author{Chong-Sun Chu${}^{1,2}$}
\author{Wu-Zhong Guo${}^{2}$ }
\affiliation{${}^1$ Department of Physics, National Tsing-Hua
  University, Hsinchu 30013, Taiwan\\
${}^2$ National Center for Theoretical Sciences, National Tsing-Hua
  University, Hsinchu 30013, Taiwan}

\date{January 16, 2017}

\begin{abstract}
We propose a new holographic dual of  conformal field theory defined
on a manifold with boundaries, i.e. BCFT. Our proposal can apply to
general boundaries and agrees with arXiv: 1105.5165 for 
the special case of a disk and
half plane. Using the new proposal of AdS/BCFT,  we successfully
obtain the expected boundary Weyl anomaly and 
the obtained boundary central charges satisfy naturally a c-like theorem holographically.
We also investigate the holographic
entanglement entropy of BCFT and find that the minimal surface must be
normal to the 
bulk spacetime boundaries when they intersect. Interestingly,
the entanglement entropy depends on the boundary conditions of BCFT
and the distance to the boundary. The entanglement wedge has an
interesting phase transition which is important for the
self-consistency of AdS/BCFT. 
\end{abstract}

\maketitle

\section{Introduction}

BCFT (Boundary Conformal Field Theory) 
is a CFT defined on a manifold $M$  with a
boundary $P$, 
and with suitable boundary conditions imposed. 
It has important
applications in string theory and condensed matter physics, 
e.g. boundary critical behavior\cite{Cardy:2004hm}. 
The  AdS/CFT correspondence  \cite{Maldacena:1997re,adscft} 
is a concrete realization of holography.  
The duality has not only opened door to previously intractable
problems in strongly 
coupled nonperturbative problems in quantum field theories (QFT), but 
has also offered  many useful insights into the fundamental properties 
of quantum gravity. 
In this regard, it is interesting to extend the AdS/CFT correspondence
to BCFT in order 
to get new handles to tackle some of the difficult problems in BCFT.
The presence of boundary in the QFT will also offer new twists in the
realization of the AdS/CFT correspondence, and should lead to a deeper
understanding of the holographic principle.

Recently, Takayanagi \cite{Takayanagi:2011zk}
proposed to extend the $d$ dimensional manifold $M$ to a $d+1$
dimensional asymptotically AdS space $N$ so that $\partial N= M\cup
Q$, where $Q$ is a $d$ dimensional manifold which satisfies $\partial
Q=\partial M=P$. 
The gravitational 
action for holographic BCFT 
is \cite{Takayanagi:2011zk,Nozaki:2012qd}
\begin{eqnarray}\label{action1}
I&=&\int_N \sqrt{G} (R-2 \Lambda) +  2\int_M
\sqrt{g}\ K + 2\int_Q \sqrt{h} (K-T)\nonumber\\
&&+2\int_P \sqrt{\sigma}\ \theta,
\end{eqnarray}
where  $\theta=\arccos(n_M \cdot n_Q)$ is the
supplementary angle between the boundaries $M$ and $Q$, and it is needed
for a well-defined variational principle for the joint $P$ 
\cite{Hayward:1993my}.
We have taken $16\pi G_N =1$.
Note that here we have 
allowed  in the action a constant  
term $T$ 
on $Q$.
 $T$  can be regarded as the holographic dual of boundary conditions of BCFT
since it affects the boundary
entropy 
(and also the boundary central charges, see  (\ref{3dBWA1},\ref{4dBWA1}) below)
which are closely related to the
boundary conditions (BC) \cite{Takayanagi:2011zk,Nozaki:2012qd}. 

A central issue in the construction of the AdS/BCFT 
is the determination of the location of $Q$ in the bulk.  
Imposing Dirichlet BC on $M$ and $P$: $\delta g_{ij}|_M=\delta
\sigma_{ab}|_P=0$, we get the variation of the on-shell action
\begin{eqnarray}\label{daction1}
\delta I&=&-\int_Q \sqrt{h}
\left(K^{\alpha\beta}-(K-T)h^{\alpha\beta}\right) 
\delta h_{\alpha\beta}.
\end{eqnarray}
Interestingly, Takayanagi \cite{Takayanagi:2011zk} proposed to impose 
Neumann BC on $Q$:
\begin{eqnarray}\label{NBC}
K_{\alpha\beta}-(K-T)h_{\alpha\beta}=0
\end{eqnarray}
to fix the 
position of $Q$.
For more general boundary conditions which break boundary conformal invariance,
\cite{Takayanagi:2011zk} proposed
to add matter fields on $Q$ and replace eq.(\ref{NBC}) by
\begin{eqnarray}\label{NBC1}
K_{\alpha\beta}-K h_{\alpha\beta}=\frac{1}{2} T^Q_{\alpha\beta},
\end{eqnarray}
where we have included $2 T h_{\alpha\beta} $ in the matter stress
tensor $ T^Q_{\alpha\beta}$.
For geometrical shape of $M$ with high symmetry such
as the case of a disk or half plane, 
(\ref{NBC}) fixes the location of $Q$ and produces many elegant
results for  BCFT \cite{Takayanagi:2011zk,Nozaki:2012qd,Fujita:2011fp}. 
However 
since 
$Q$ is of co-dimension one and its shape is determined by a single 
embedding function,
(\ref{NBC}) gives too many constraints  and there is no solution
in a given metric such as $AdS$ generally. 
On the other hand, of course, there should exist
well-defined BCFT with general boundaries.  
As motivated in \cite{Takayanagi:2011zk,Nozaki:2012qd},
 \eq{NBC} and \eq{NBC1}  
are natural from the 
viewpoint
of braneworld
scenario. However from a practical point
of view,  it is
not entirely satisfactory since one
has a large freedom to choose the matter fields as long as they
satisfy various energy conditions. As a result, it seems one can put the
boundary $Q$ at almost any position as one likes. 
Besides, it is unappealing that
the holographic dual depends on the details of matters on $Q$.
Finally, although eq.(\ref{NBC1}) could have solutions by
tuning the 
matters, it is 
actually too strong since one can show that \cite{parallelpaper} it 
always makes vanishing some of the central charges 
in the boundary Weyl anomaly.  In this letter, we
propose a new holographic
dual of BCFT with $Q$ determined by a new  condition \eq{mixedBCN2}.
This condition is consistent and provides a unified treatment to 
general shapes of $P$. 
Besides, as we will show below, it yields the expected boundary
contributions to Weyl anomaly.

\section{New Proposal for Holographic BCFT}
 Instead of imposing the
Neumann 
BC (\ref{NBC}), we propose to impose on $Q$ the mixed 
BCs,  $\Pi_{-\alpha\beta}^{\ \ \  \alpha'\beta'}\delta
h_{\alpha'\beta'}=0$ and
\begin{eqnarray}\label{mixedBCN}
(K^{\alpha\beta}-(K-T)h^{\alpha\beta})\Pi_{+\alpha\beta}^{\ \ \ \alpha'\beta'}=0.
\end{eqnarray}
Here $\Pi_\pm$ are projection operators satisfying
$\Pi_{+\alpha\beta}^{\ \ \  \alpha'\beta'}+\Pi_{-\alpha\beta}^{\ \ \  \alpha'\beta'}
=\delta_{\alpha}^{\alpha'}\delta_{\beta}^{\beta'}$
and $\Pi_{\pm \alpha\beta}^{\ \ \  \alpha'\beta'}
\Pi_{\pm  \alpha'\beta'}^{\ \ \  \alpha_1\beta_1}
=\Pi_{\pm  \alpha\beta}^{\ \ \  \alpha_1\beta_1}$. 
Since we could impose at
most one condition to fix the location of the co-dimension one surface
$Q$, we require $\Pi_+$ to be of the form $\Pi_{+\alpha\beta}^{\ \ \  \alpha'\beta'}=
A_{\alpha\beta}B^{ \alpha'\beta'}$. $\Pi_{+}\Pi_{+}=\Pi_{+}$ then implies 
${\rm tr} AB^T=1$. 
The mixed boundary condition \eq{mixedBCN} becomes
\begin{eqnarray}\label{mixedBCN1}
(K^{\alpha\beta}-(K-T)h^{\alpha\beta})A_{\alpha\beta}=0,
\end{eqnarray}
where $A_{\alpha\beta}$ are 
to be determined. 
It is natural to require that
eq. (\ref{mixedBCN1}) to be linear in $K$ so that it is a second order
differential equation for the embedding. 
In this paper we propose the choice
$A_{\alpha\beta}=h_{\alpha\beta}$. We will show below that 
there are problems with
the other choices such as
\be \label{otherA}
A_{\alpha\beta}=\lambda_1 h_{\alpha\beta}+ \lambda_2
K_{\alpha\beta}+\lambda_3 R_{\alpha\beta}+ \cdots,
\;\;\;
\lambda_1, \lambda_2 \neq 0, \;\;\; 
\ee

To sum up, we propose to use the traceless condition 
\begin{eqnarray}\label{mixedBCN2}
T_{BY}{}^{\alpha}_{\ \alpha}=2(1-d)K+2dT=0
\end{eqnarray}
to determine the boundary $Q$. Here
$T_{\rm BY}{}_{\alpha\beta}=2K_{\alpha\beta}-2(K-T)h_{\alpha\beta}$ 
is the Brown-York stress tensor on $Q$.
In general, it could also depend on the intrinsic curvatures which we
will treat in the paper \cite{parallelpaper}. 
A few remarks on \eq{mixedBCN2} are in order.
{\it 1.} It is worth noting that the junction 
condition for a thin shell with spacetime on both sides 
is 
also given by (\ref{NBC1}) \cite{Hayward:1993my}. However, here
$Q$ is the
boundary of spacetime and not a thin shell,
so there is no need to consider
the junction condition. 
{\it 2.} For the same reason,  it is expected that $Q$ has
no back-reaction on the geometry just as the boundary $M$. 
{\it 3.} Eq. (\ref{mixedBCN2}) implies that $Q$ is a
constant mean curvature surface,
which is also of great interests in both
mathematics and physics just as 
the minimal surface. 
{\it 4.} \eq{mixedBCN2}  reduces to the proposal by
\cite{Takayanagi:2011zk} for a disk and half-plane. And it can
reproduce all the results in
\cite{Takayanagi:2011zk,Nozaki:2012qd,Fujita:2011fp}. 
{\it 5.} 
Eq. \eq{mixedBCN2} is a purely geometric equation and
has solutions for arbitrary shapes of boundaries and arbitrary bulk
metrics.
{\it 6.} Very importantly,
our proposal 
gives non-trivial boundary Weyl anomaly, which solves the difficulty met in
\cite{Takayanagi:2011zk,Nozaki:2012qd}.
In fact one can show that \cite{parallelpaper} 
the proposal (\ref{NBC1}) 
is too restrictive
and always yields $c_2=b_1=0$ 
in (\ref{3dBWA},\ref{4dBWA}) \cite{parallelpaper}.

Let us recall that in the presence of boundary,  Weyl anomaly 
of CFT generally pick up a boundary contribution
$\left<T^a_a\right>_P$ in addition to the usual bulk term
$\left<T^i_i\right>_M$, i.e.
$ \left< T^i_i \right>=\left<T^i_i\right>_M
+\delta(x_{\perp})\left<T^a_a\right>_P$,
where $\delta(x_{\perp})$ is a delta function with support on the
boundary $P$. Our proposal yields
the expected boundary Weyl anomaly for 3d and 4d BCFT 
\cite{Herzog:2015ioa,Fursaev:2015wpa,Solodukhin:2015eca}: 
\begin{eqnarray}\label{3dBWA}
&&\left<T^a_a\right>_P= c_1 \mathcal{R}+ c_2 \text{Tr} \bar{k}^2, \quad
  d=3, \\
&&\left<T^a_a\right>_P=\frac{a}{16\pi^2} E_4^{\rm bdy}+ b_1 \text{Tr}\bar{k}^3 + b_2
  C^{ac}_{\ \  b c} \bar{k}_{\ a}^b,\quad
d=4, \;\;\;\;\;\; \label{4dBWA}
\end{eqnarray}
where $c_1, c_2, b_1,b_2$ are boundary central charges, $a=2\pi^2$ is
the bulk central charge for 4d CFTs dual to Einstein gravity.  
$\mathcal{R}$ and $\bar{k}_{ab}$ are the intrinsic
curvature and the 
traceless part of the extrinsic curvature of $P$,  
$C_{abcd}$ is the pull back of the Weyl
tensor 
of $M$ to $P$, and
\begin{eqnarray}\label{E4bry}
 E^{\rm bdy}_4=4\Big(2 \text{Tr} (k\mathcal{R})-k\mathcal{R}+\frac{2}{3}
 \text{Tr} k^3- k \text{Tr} k^2+\frac{1}{3}k^3\Big) \;\;\;\;\;\;\;
\end{eqnarray}
is the boundary terms of the Euler density
$E_4= R_{ijkl} R^{ijkl} - 4R_{ij}R^{ij} + R^2$
in order to preserve the topological invariance. 
Since $Q$ is not a minimal surface in our case, our results (\ref{3dBWA1},\ref{4dBWA1}) are non-trivial generalizations of the Graham-Witten anomaly \cite{Graham:1999pm} for the submanifold.

\section{Holographic Boundary Weyl Anomaly}

\textbf{Action method.} Applying the method of
\cite{Henningson:1998gx}, one can derive the Weyl anomaly (including
the boundary Weyl anomaly \cite{Nozaki:2012qd}) as the logarithmic
divergent term of the gravitational action. For our purpose, we focus
only on the boundary contributions to Weyl anomaly below.

Consider the asymptotically AdS metric in the Fefferman-Graham gauge 
\begin{eqnarray}\label{AdSmetric}
ds^2=\frac{dz^2+g_{ij}dx^idx^j}{z^2}, 
\end{eqnarray}
where $g_{ij}=g^{(0)}_{ij}+z^2 g^{(1)}_{ij}+ \cdots$, $g^{(0)}_{ij}$ is
the metric of BCFT on $M$ , $g^{(1)}_{ij}$ can be fixed by the PBH
(Penrose-Brown-Henneaux) transformation \cite{Imbimbo:1999bj}
\begin{eqnarray}\label{AdSmetric1}
g^{(1)}_{ij}=-\frac{1}{d-2}(R^{(0)}_{ij}-\frac{R^{(0)}}{2(d-1)}g^{(0)}_{ij}). 
\end{eqnarray}
Note that the curvatures in our notation differ from those of
\cite{Imbimbo:1999bj} by a minus sign.
Without loss of generality, 
we choose the Gauss normal coordinates for the metric $g^{(0)}_{ij}$:
\begin{eqnarray}\label{BCFTmetric}
ds_0^2=dx^2+(\sigma_{ab}+2x k_{ab}+ x^2 q_{ab}+ \cdots)dy^a dy^b,
\end{eqnarray}
where $P$ is located at $x=0$
and $y^a$ are the coordinates along $P$. The bulk
boundary $Q$ is given by $x=X(z,y)$. Expanding it in $z$, 
\begin{eqnarray}\label{Qsuface}
x=a_1 z+ a_2 z^2+ \cdots + ( b_{d+1} \ln z+ a_{d+1}) z^{d+1}+ \cdots,
\;\;\;\;\;\;\;
\end{eqnarray}
where the coefficients $a$'s and $b$'s are functions of $y$.
Substituting eqs.(\ref{AdSmetric}- \ref{Qsuface}) into the boundary
condition eq.(\ref{mixedBCN2}), we obtain that 
\begin{eqnarray}\label{uniTa}
T=(d-1)\tanh\rho, \; a_1= \sinh\rho , \;
a_2=-\frac{\cosh^2\rho \,{\rm Tr}k}{2(d-1)}, \;\;\;\;\;\;\;
\end{eqnarray} 
where we have re-parametrized the constant $T$.  
It is
worth noting that the other choices \eq{otherA} of $A_{\alpha\beta}$ 
gives the same $T, a_1, a_2$  but different $a_3,
a_4, \cdots$. In other words, the results 
(\ref{uniTa}) are independent of the
choices of $A_{\alpha\beta}$ in the boundary condition
(\ref{mixedBCN1})
\cite{parallelpaper}. 
In fact since 
$K^{\alpha}_{\beta}=\frac{a_1}{\sqrt{1+a_1^2}}\delta^{\alpha}_{\beta}+O(z)$,
one obtains from (\ref{mixedBCN1}) that
$ (1-d)\frac{a_1}{\sqrt{1+a_1^2}}+T=0$ 
as long as $A^{\alpha}_{\alpha} \neq 0$. This gives
the first two terms in (\ref{uniTa}). 
As for the coefficient $a_2$, according to \cite{Schwimmer:2008yh}, the
embedding function eq.(\ref{Qsuface}) is highly constrained by the
asymptotic symmetry of AdS, and it can be fixed by PBH transformations
up to some conformal tensors. Adapting the method of
\cite{Schwimmer:2008yh} to the present case, 
one can indeed prove the universality of
$a_2$ in the Gauss normal
coordinates \cite{parallelpaper}. 
In this way, we obtain $a_2=-\frac{\cosh^2\rho\text{Tr}k}{2(d-1)}$, which
agrees with the result obtained in \cite{Schwimmer:2008yh} for the
special case of $a_{\rm odd}=\rho=0$.

Now we are ready to derive the boundary Weyl anomaly. For simplicity,
we focus on the case of 3d BCFT and 4d BCFT.  Substituting
eqs.(\ref{AdSmetric}-\ref{uniTa}) into the action (\ref{action1}) and
selecting the logarithmic divergent terms after the integral along $x$
and $z$, we can obtain the boundary Weyl anomaly.  
We note that 
$I_M$ and $I_P$ do not contribute to the logarithmic divergent term
in the action since they have at most singularities in powers of $z^{-1}$  
but there is
no integration alone $z$, thus there is no way for them 
to produce $\log z$ terms. 
We also note that only $a_2$ appears in the final results. The terms
including $a_3$ and $a_4$ automatically cancel each other out. This is
also the case for the holographic Weyl anomaly and universal terms of
entanglement entropy for 4d and 6d CFTs \cite{Miao:2013nfa,Miao:2015iba}. 
After some calculations, we obtain
the boundary Weyl anomaly for 3d and 4d BCFT as
\begin{eqnarray}\label{3dBWA1}
&&\left<T^a_a\right>_P= \sinh \rho\  \mathcal{R}-\sinh
  \rho\  \text{Tr} \bar{k}^2,\\
&&\left<T^a_a\right>_P= \frac{1}{8} E_4^{\rm bdy}+
  \left(\cosh(2\rho)-\frac{1}{3}\right) \text{Tr} \bar{k}^3 \nonumber\\
&& \ \ \ \ \ \ \ \ \ \ \ \ \ \ \ \  \ \ \ \ \ \ 
-\cosh(2\rho) C^{ac}_{\ \ \ b c} \bar{k}_{\ a}^b, \label{4dBWA1} 
  \end{eqnarray}
which takes the expected form \eq{3dBWA}, \eq{4dBWA}. It is
remarkable that the coefficient of $ E_4^{\rm bdy}$ takes the correct
value to  preserve the topological invariance of $E_4$. This is a
non-trivial check of our results. Besides, the 
boundary charges $c_1, b_1$ in (\ref{3dBWA}, \ref{4dBWA})
are expected to satisfy a c-like theorem \cite{Nozaki:2012qd,Jensen:2015swa,Huang:2016rol}. 
As was shown in \cite{Takayanagi:2011zk,Fujita:2011fp}, null energy condition on
$Q$ implies $\rho$ decreases along RG flow. It is also true for us. As a
result, 
eqs.(\ref{3dBWA1}, \ref{4dBWA1}) indeed obey the c-theorem for
boundary charges.
This is also
a support for our results.  Most importantly, our confidence is based
on the above universal derivations, i.e., we do not make any
assumption about $A_{\alpha\beta}$ in the boundary condition
(\ref{mixedBCN1}).

We remark that 
based on the results of free CFTs \cite{Fursaev:2015wpa} and the
variational principle, it 
has been suggested that the coefficient of $Ck$
in (\ref{4dBWA1}) is universal for all 4d BCFTs
\cite{Solodukhin:2015eca}. Here we provide 
evidence, based on holography, against this suggestion: 
our results agree with the suggestion of
\cite{Solodukhin:2015eca} for the trivial case $\rho=0$, 
while disagree generally.  As argued in \cite{Huang:2016rol}, the
proposal of \cite{Solodukhin:2015eca} is suspicious. It
means that there could be no independent boundary central charge
related to the Weyl invariant
$\sqrt{\sigma} C^{ac}_{\ \ \ b c} \bar{k}_{\ a}^b$. However, in general, every
Weyl invariant should correspond to an independent central charge,
such as the case for 2d, 4d and 6d CFTs.
Besides, we notice
that the law obeyed by free CFTs usually does not apply to strongly
coupled CFTs. See
\cite{Dong:2016wcf,Lee:2014zaa,Hung:2014npa,Chu:2016tps} for 
examples. 
 
In this subsection, we have proved that, by
using the method of \cite{Henningson:1998gx}, all the possible
boundaries $Q$ allowed by (\ref{mixedBCN1}) produce the same
boundary Weyl anomaly for 3d and 4d BCFT. Thus this method cannot distinguish the proposal (\ref{mixedBCN2}) from
the other choices \eq{otherA}.

\textbf{Stress-tensor method.} 
To resolve the above ambiguity, let us use the holographic stress
tensor \cite{Balasubramanian:1999re} to study the boundary Weyl
anomaly as this method needs the information of $(a_3, a_4,\cdots)$ which can  distinguish different choices of boundary conditions (\ref{mixedBCN1}).
For simplicity, we focus on the case of 3d BCFT.

The first step of method \cite{Balasubramanian:1999re} is to find a
finite action by adding suitable covariant counterterms. We obtain
\begin{eqnarray}\label{regularizedaction1}
I_{\rm ren}&=&\int_N \sqrt{G} (R-2 \Lambda) 
+ 2\int_M \sqrt{g}(K-2-\frac{1}{2}R_M )\nonumber\\
&+& 2\int_Q \sqrt{h}( K-T)+2\int_P  \sqrt{\sigma}(\theta
-\theta_0- K_M ),\;\;\;\;\;
\end{eqnarray}
where we have included on $M$ the usual counterterms in holographic
renormalization \cite{Balasubramanian:1999re,deHaro:2000vlm},
$\theta_0=\theta(z=0)$ is a constant \cite{Nozaki:2012qd} and $K_M$, the
extrinsic curvature of $P$, is
the Gibbons-Hawking-York term for $R_M$ on $M$. Notice that there is
no freedom to add other counterterms, except for some finite terms which
are irrelevant to Weyl anomaly. For example, we may add
terms like 
$\sqrt{\sigma}{\cal R}$ and $\sqrt{\sigma}K_M^2$ to $I_P$. However, these
terms are invariant under constant Weyl transformations. Thus they do
not contribute to the boundary Weyl Anomaly. In conclusion, the
renormalized action (\ref{regularizedaction1}) is unique up to some
irrelevant finite counterterms.

 From the 
renormalized action, it is straightly to derive the
 Brown-York stress tensor on $P$
\begin{eqnarray}\label{regularizedstresstensor1}
B_{ab}=2(K_{M ab}-K_M \sigma_{ab})+2(\theta -\theta_0)\sigma_{ab}
\end{eqnarray}
In the sprint of \cite{Nozaki:2012qd,Balasubramanian:1999re,deHaro:2000vlm}, the
boundary Weyl anomaly is given by
\begin{eqnarray}\label{HBWAstress1}
\left<T^a_a\right>_P=\lim_{z\to 0} \frac{B^a_a}{z^2}=\lim_{z\to 0}
\frac{4(\theta-\theta_0)-2 K_M}{z^2},
\end{eqnarray}
where 
$\theta= \cos^{-1} \frac{x'}{\sqrt{g^{xx}+x'^2}}+O(z^3)$, $\theta_0=
\cos^{-1}(\tanh \rho)$ and 
$K_M=z\frac{\partial_x (\sqrt{g}\sqrt{g^{xx}})}{\sqrt{g}}+O(z^3)$. 
Substituting
eqs.(\ref{AdSmetric}-\ref{uniTa}) into \eq{HBWAstress1}, we
get
\begin{eqnarray}\label{HWAmstress1}
 \left<T^a_a\right>_P&=&-\frac{{\rm sech}^2 \rho }{4}  \Big[48
   a_3+\sinh 3 \rho  \left(2 q- 3 k^2-4 \text{Tr}\bar{k}^2\right)\nonumber\\
&&+\sinh \rho  \left(2 \mathcal{R}+6 q-6 k^2-6 \text{Tr}\bar{k}^2\right)\Big],
\end{eqnarray}
where $q$ is the trace of $q_{ab}$. This gives the correct 
boundary Weyl anomaly (\ref{3dBWA1}) if and only if 
\begin{eqnarray}\label{a3}
 a_3 &=&\frac{1}{48} \sinh\rho \Big[
\cosh (2 \rho ) (-2 \mathcal{R}-4 q+k^2+10
 \text{Tr}k^2)\nonumber\\
 &&-4\mathcal{R}-8 q+3k^2+12 \text{Tr}k^2\Big],
\end{eqnarray}
which is just the solution to our proposed boundary condition
(\ref{mixedBCN2}).
One can check that the other choices 
\eq{otherA}  give different
$a_3$ and thus can be excluded. Following the same approach, we can
also derive the boundary Weyl anomaly for 4d BCFT \cite{parallelpaper},
which agrees with the correct result \eq{4dBWA1} 
iff $a_3$ and $a_4$ are
given by the solutions to eq.(\ref{mixedBCN2}). This is a very strong
support to the boundary condition (\ref{mixedBCN2}) we proposed.

\section{Holographic Entanglement Entropy}

Following \cite{Ryu:2006bv,Lewkowycz:2013nqa}, it is not difficult to
derive the holographic entanglement entropy 
for
BCFT, which is also
given by the area of  minimal surface 
 \begin{eqnarray}\label{HEE}
S_A=\frac{\text{Area}(\gamma_A)}{4 G_N},
\end{eqnarray}
where $A$ is a 
subsystem on $M$, and $\gamma_A$ denotes the minimal
surface which ends on $\partial A$.   What is new for BCFT is that the
minimal surface could also end on the bulk boundary $Q$, when the
subsystem $A$  is close to the boundary $P$. See Fig.1 for example.

We could keep the endpoints of extreme
surfaces $\gamma_A'$ freely on $Q$, and select the one with minimal
 area as $\gamma_A$. It follows that
$\gamma_A$ is orthogonal to the boundary $Q$ when they
intersect
 \begin{eqnarray}\label{normalAQ}
n^a_{\gamma_A} \cdot n_Q|_{\gamma_A\cap Q}=0.
\end{eqnarray}
Here $n_Q$ is the normal vector of $Q$ and 
$n^a_{\gamma_A}$ are the two independent normal vectors of
$\gamma_A$. Another way to see this is that,
otherwise there will arise problems in the holographic
 derivations of entanglement entropy by using the replica trick. In
 the replica method, one considers the $n$-fold cover $M_n$ of $M$ and
 then extends it to the bulk as $N_n$. It is important that $N_n$ is a
 smooth bulk solution. As a result, Einstein equation should be
 smooth on 
surface $\gamma_A$.
Now the metric near  $\gamma_A$
is given by \cite{Lewkowycz:2013nqa}
 \begin{eqnarray}\label{minimalsurface}
ds^2=\frac{1}{r^{2\varepsilon}}(dr^2+r^2 d\tau^2)+
\left(g_{ij}+2\mathcal{K}_{aij}x^a+ O(r^2) \right)dy^i
dy^j, \nonumber
\end{eqnarray}
where 
$\varepsilon \equiv 1-\frac{1}{n}$, $r$ is coordinate normal to the
surface, $\tau\sim \tau+2\pi n$ is the Euclidean time, $y^i$ are coordinates along the surface,
$x^a =(r \cos \tau, r \sin \tau)$ and
$\mathcal{K}_{aij}$ are the two extrinsic
curvature tensors. Going to complex
coordinates $z=r e^{i\tau}$, the $zz$ component of Einstein equations
 \begin{eqnarray}\label{Eineq}
R_{zz}=-\mathcal{K}_z \frac{\varepsilon}{z}+ \cdots
\end{eqnarray}
is divergent unless the trace of extrinsic curvatures vanish
$\mathcal{K}_a=0$. This gives 
the condition for a minimal
surface \cite{Lewkowycz:2013nqa}.  
Labeling the boundary $Q$ by $f(z,\bar{z},y)=0$, we obtain
the extrinsic curvature of $Q$ as
 \begin{eqnarray}\label{BCregular}
K\sim \varepsilon \ \partial_z f \partial_{\bar{z}} f (\frac{ \partial_z
  f }{\bar{z}}+\frac{ \partial_{\bar{z}} f }{z} )+ \cdots . 
\end{eqnarray}
So the boundary condition (\ref{mixedBCN2}) is smooth only if
$\partial_z f|_{\gamma_A\cap Q} = \partial_{\bar{z}} f|_{\gamma_A\cap
  Q}=0 $, which is exactly the
orthogonal condition (\ref{normalAQ}). 
As a summary, the holographic entanglement entropy for BCFT is 
given by RT
formula (\ref{HEE}) together with the orthogonal condition
(\ref{normalAQ}).

\section{Boundary Effects on Entanglement}

Let us take a simple example to illustrate the boundary 
effects on entanglement entropy. Consider Poincare metric of $AdS_3$
$ds^2=(dz^2+dx^2-dt^2)/z^2$,
where $P$ is at $x=0$. For simplicity, we focus on  $T=\tanh \rho \ge
0$ below. Solving eq.(\ref{mixedBCN2}) for $Q$, 
we get $x= \sinh(\rho)  z$. We choose $A$ as an
interval with two endpoints at $x=d$ and $x=d+2l$. Due to the presence
of boundary, there are now two kinds of minimal surfaces, one ends on
$Q$ and the other one does not. It depends on the distance $d$ that
which one has smaller area. 
 From eqs.(\ref{HEE},\ref{normalAQ}), we obtain
\be \label{SA}
S_A= \begin{cases}
\frac{1}{2 G_N}\log (\frac{2l}{\epsilon }), &  
d\ge d_c,\\
\frac{\rho}{2 G_N}+\frac{1}{4 G_N}\log 
\Big( \frac{4d(d+2l)}
{\epsilon ^2}
\Big), 
& d \le
d_c,
 \end{cases}
\ee
where $d_c=l\sqrt{e^{-2 \rho }+1}-l$ is the critical distance
and the parameter $\rho$ can be regarded as the holographic dual of the
boundary condition of BCFT.  It is
remarkable that entanglement entropy (\ref{SA}) depends on the
distance $d$ and boundary condition $\rho$ when it is close enough to
the boundary. This behavior is expected  from the viewpoint of
BCFT since it has also been found that  
the correlation functions depend on the distance to the
boundary \cite{McAvity:1993ue}.

To extract the effects of boundary on the entanglement entropy, 
let us define the following quantity when 
$A$ does not intersect the boundary $P$:
\begin{eqnarray}\label{muinformation}
I_A \equiv S^{\rm CFT}_A-S^{\rm BCFT}_A.
\end{eqnarray}
The complementary situation where the entangling surface intersects the boundary
$P$ is discussed in \cite{Takayanagi:2011zk}.
Here in \eq{muinformation} 
$S^{\rm CFT}_A$ is the entanglement entropy when the boundary
disappears or is at infinity. 
In the holographic language, it is given
by the area of minimal surface that does not end on $Q$. Thus,
$S^{\rm CFT}_A$ is equal to or bigger than $S^{\rm BCFT}_A$ and $I_A$ is
always 
non-negative. 
It is expected that boundary does not affect the
divergent parts of entanglement entropy when $A\cap P=0$, so all the
divergence cancel in eq.(\ref{muinformation}). As a result, $I_A$ is
not only  non-negative but also finite. 
Physically, $I_A$ measures the decrease in the entanglement of the subsystem
$A$ with the environment when a boundary is introduced.
For the example discussed above, we
find
\begin{eqnarray}\label{IA}
I_A=\begin{cases}
0, & d \ge d_c\\
\frac{1}{4 G}\log (\frac{l^2}{d (d+2 l)})-\frac{\rho }{2 G}, & 0<d < d_c,
 \end{cases}
\end{eqnarray}
which is indeed both non-negative and finite. 
Note that $I_A$ depends both on the distance from the
boundary and the boundary condition when $d<d_c$, but  becomes
independent of them when $d \ge d_c$. This represents some kind of
phase transition. 
It is also intriguing to note that, in this simple
example,  $I_A$ is just one half of the mutual information between $A$
and its mirror image $A'$, so it must be non-negative and finite. See Fig1 for
example. 
\begin{figure}[t]
\centering
\includegraphics[width=8cm]{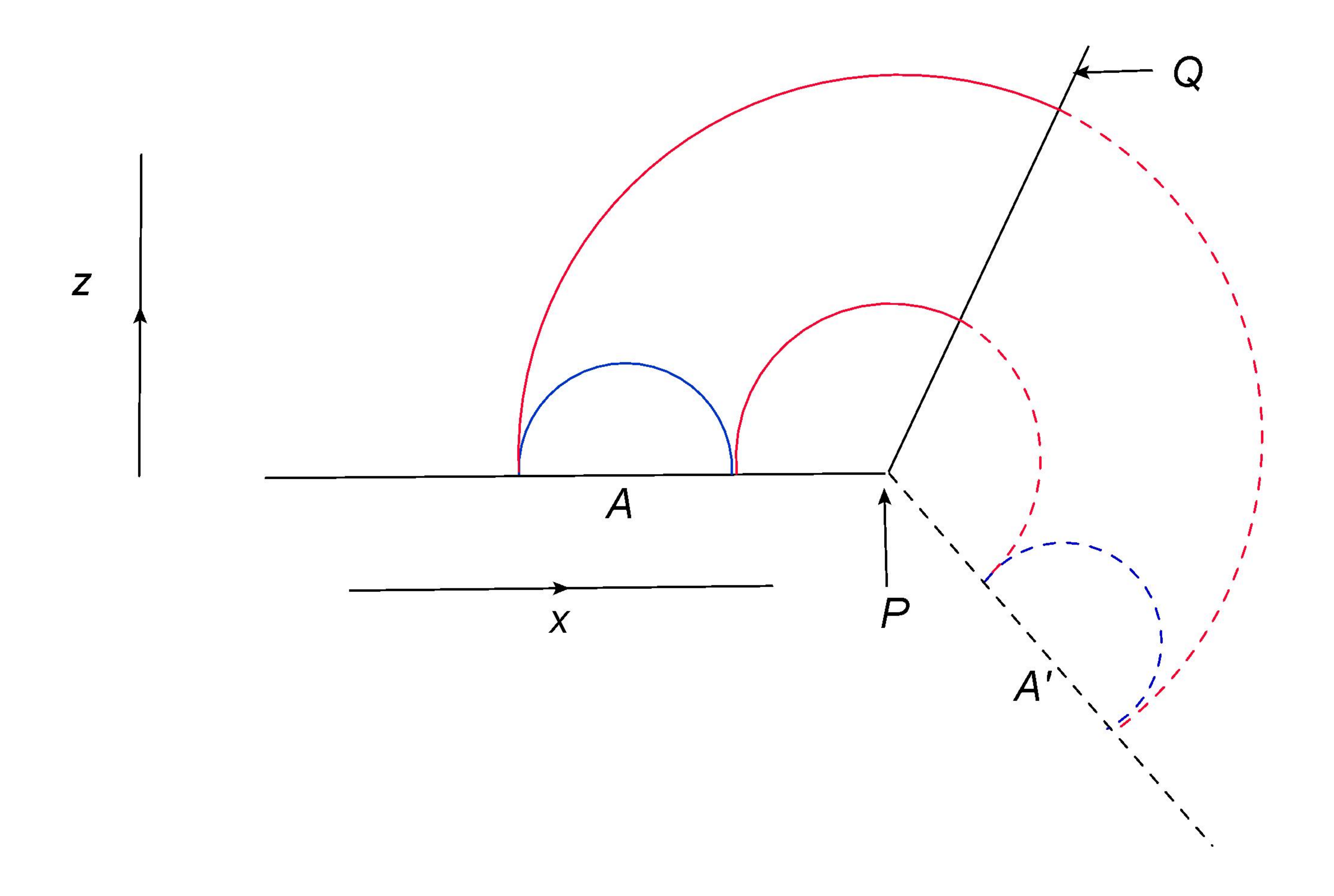}
\caption{Subsystem $A$ and its mirror image $A'$}
\end{figure}

\begin{figure}[t]
\centering
\includegraphics[width=8cm]{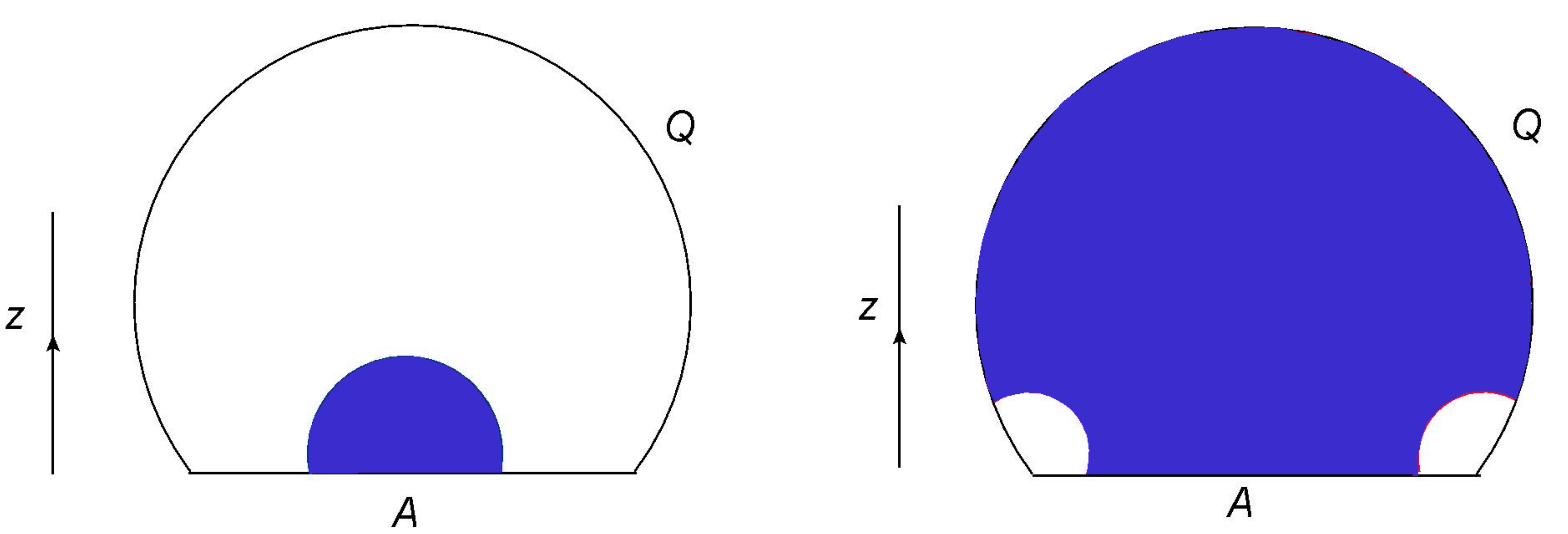}
\caption{Entanglement wedge for small $A$ and large $A$}
\end{figure}

\section{Entanglement Wedge}

According to \cite{Jafferis:2015del,Dong:2016eik}, a sub-region $A$ on
the AdS boundary is dual to 
an entanglement wedge $\mathcal{E}_A$ in
the bulk where all the bulk operators within $\mathcal{E}_A$ can
be reconstructed by using only the operators of $A$. The entanglement
wedge is defined as the bulk domain of dependence of any achronal bulk
surface between the minimal surface
$\gamma_A$ and the  subsystem $A$.

It is interesting to study the entanglement wedge in 
AdS/BCFT. For simplicity, let us focus on the static spacetime and
constant time slice.
 A key observation is that 
entanglement wedge 
behaves a
phase transition and becomes much larger than that within AdS/CFT,
when $A$ is increasing and approaching to the boundary.  See Fig.2 for
example. This  phase transition is important for the self-consistency
of holographic BCFT. If there is no phase transition, then 
$\mathcal{E}_A$ is always given by the first kind ( left hand side
of Fig.2). When $A$ fills with the whole boundary $M$ and $P$, there
are still large space left outside the entanglement wedge, which means
there are operators in the bulk cannot be reconstructed by all the
operators on the boundary. Thanks to the phase transition,
$\mathcal{E}_A$
for large A is given by the second kind ( right
hand side of Fig.2). As a result all the bulk operators can be
reconstructed by using the boundary operators.

\section{Conclusions and Discussions}

In this letter, we propose a new holographic dual of BCFT, which
can accommodate all possible shapes of the boundary $P$ 
in a unified prescription.  The key idea is
to impose the mixed boundary condition \eq{mixedBCN2} so that there is only one
constraint for the co-dimension one boundary $Q$. 
In general there could be 
more than one self-consistent boundary 
conditions for a theory \cite{Song:2016pwx}, so the
proposals of \cite{Takayanagi:2011zk} and ours have no
contradiction in principle. However, the proposal of
\cite{Takayanagi:2011zk} is too 
restrictive to include the general BCFT. 
The main advantage of our proposal is that we can deal
with all shapes of the boundary 
$P$ easily. 
It is appealing that the
bulk boundary $Q$ is given by a constant mean curvature surface, which
is a natural generalization of the minimal surface.

Applying the new AdS/BCFT, we obtain the expected boundary Weyl
anomaly
and the obtained boundary central charges satisfy naturally a c-like
theorem holographically.
As a by-product, we give a holographic disproof of the 
proposal
\cite{Solodukhin:2015eca} and clarify that the validity of the $S_{RE}=S_{EE}$ conjecture
\cite{Herzog:2016kno} based on \cite{Solodukhin:2015eca} sensitively depends on the 
boundary conditions of non-free BCFT.
Besides, we find the holographic entanglement entropy is given by 
the RT formula 
together with the condition
that the minimal surface must be orthogonal to $Q$ if they intersect.
The presence of boundaries lead to many interesting effects, 
e.g. phase transition of the entanglement wedge.
Of course, many things are
left to be explored, for instance, the edge modes
\cite{Donnelly:2014fua,Huang:2014pfa}, the shape dependence of
entanglement \cite{Bueno:2015rda,Mezei:2014zla}, the
applications to condensed matter and
the relation between BCFT and quantum information
\cite{Numasawa:2016emc}.  
Finally, it is straightforward to
generalize our work to Lovelock gravity, higher dimensions and general
boundary conditions.

\section*{Acknowledgements}
We would like to thank 
X. Dong, L.Y. Hung and F.L. Lin for useful discussions and comments.
 This work is supported in part by the National
Center of Theoretical Science (NCTS) and the grant MOST
105-2811-M-007-021 of the Ministry of
Science and Technology of Taiwan.



\begin{thebibliography}{99}
\bibitem{parallelpaper}
Please see the  
companion paper for the proof of the universality of $a_2$ by 
using the PBH transformations and the details
of holographic renormalization for 4d BCFT, where the boundary Weyl
anomaly is derived.
\end{thebibliography}
\end{document}